# Applying the Perturbative Integral in Aeromaneuvers around Mars to Calculate the Cost

Jhonathan O. Murcia Piñeros*
Corresponding author: jhonathan.pineros@unifesp.br
Institute of Science and Technology ICT-UNIFESP, São José dos Campos, (SP), Brazil.
ORCID: 0000-0002-7013-6515

Antônio F. Bertachini de Almeida Prado
National Institute for Space Research INPE, (Graduate Division - DIPGR), São José dos Campos, (SP), Brazil.
ORCID: 0000-0002-7966-3231

Walter Abrahão dos Santos
National Institute for Space Research INPE, (Small Satellite Division - DIPST), São José dos Campos, (SP), Brazil.
ORCID: 0000-0002-6516-6488

Rodolpho Vilhena de Moraes
Institute of Science and Technology ICT-UNIFESP, São José dos Campos, (SP), Brazil.
ORCID: 0000-0003-1289-8332

**Abstract**
The perturbative integral method was applied to quantify the contribution of external forces during a specific interval of time in trajectories of spacecrafts around asteroids and under the Luni-solar influence. However, this method has not been used to quantify the contributions of drag in aerocapture and aerobraking, for this reason, the planet Mars is select to apply the method during an aerogravity-assisted maneuver. Several trajectories are analyzed making use of a drag device, with area to mass ratios from 0.0 to 20.0 $m^2$/kg, simulating solar sails or de-orbit devices. The mathematical model consists in the restricted three-body problem. The use of this maneuver makes possible to obtain variations of energy in the trajectory replacing expensive maneuvers based in fuel consumption. To observe the effects of the maneuvers, different values of pericenter velocity and altitude were selected for prograde and retrograde orbits. The innovation of this research is the application of an integral method to quantify the delta-V of the aero gravity maneuver, comparing the cost of the maneuver with the traditional methods of space propulsion. The results allow to identify orbits with conditions to capture, and the perturbative maps show the velocity variations in specific regions, given by the drag.

1. Introduction

A method to quantify the influence of orbital perturbations called Perturbative Integral (PI) was presented in Refs. [1–3]. The PI quantifies the contribution of each acceleration involved in the spacecraft trajectory. It was implemented to find the less perturbed orbits around the Earth and under Lunisolar influence [2]; was applied to map orbits around asteroids [3] and to analyze the effect of the terms of the expansion of the gravity field of the Earth [1]. Other applications include the quantification of perturbative forces in harmonic and duffing oscillators and measurements of the perturbations in satellites [4]. Those ideas were generalized, and four types of PI were described in the literature to calculate perturbation maps around asteroids [5] and to quantify the effects of the accelerations during a Powered Aero-Gravity Assisted Maneuver (PAGAM) [6]. The results showed that the PI measures the evolution of the perturbations, indicating the Delta-V (DV) given by those forces and so predicting the orbital changes they made in the trajectory of the spacecraft. An important improvement appeared in the literature when studying the irregular shape of the Earth [7], which made an important step in this type of integral index and developed a version that is much more accurate in several types of applications, when there are forces in opposed directions involved in the trajectory.

As was presented, the PI was applied in several problems in orbital dynamics, however, has not been used to analyze aeromaneuvers, with focus on aerocapture, which is useful for future interplanetary flights. For example, the PI could be applied to quantify the contribution of the aerodynamic forces during a final approach in planets like Venus, Mars, and Jupiter, which present a significant atmospheric density. Conceptual missions are projected to include aerocapture and aerobraking maneuvers, instead of traditional propulsion systems, reducing the costs of the mission [8-12].

Successful missions to Mars, like the InSight (2018), showed that interest in the exploration of the planet continues to grow. A highlight of InSight was the transportation of two small spacecraft: MarCO A and MarCO B, the first interplanetary CubeSats. The sizes of the SmallSats restricted the use of traditional propulsion systems, due to the limited mass and volume. This scenario allows the exploration of alternative techniques as aeromaneuvers, which are possible due to the implementation of drag devices, a low-cost technology with a higher level of maturity, compared to the small propulsion systems for NanoSats.

Considering the use of small spacecraft in space exploration missions, the maturity of the technology, and the future missions to Mars, the present paper contributes to this topic by applying the PI technique to calculate the contribution of the drag during aeromaneuvers. The idea is to measure the variation of velocity given by drag during the whole trajectory of the spacecraft, so the savings in propulsion systems can be measured and expressed by the PI scalar index.

Due to the close approach, the resulting trajectory could be a Gravity Assisted Maneuver (GAM), or swing-by, and, with the pericenter located inside the atmosphere, in an Aero-Gravity Assisted Maneuver (AGAM) [13-16]. If the conditions are optimal, the approach can end in aerocapture. Different studies about the aerocapture have been discussed in the scientific literature [8-12, 17-20], showing the DV savings due to the aerodynamic effect of the atmosphere of Mars.

The next section describes the mathematical model of the trajectory and the mapping techniques of the PI used to quantify the contribution of drag as a function of the spacecraft's Area-to-Mass ratio (*A/m*). Section III presents the methodology, followed by the discussion of the results obtained and the conclusions.

## 2. Modeling the trajectory

The mathematical model selected for the simulations is the Circular Restricted Three-Body Problem (CRTBP), a model that was implemented with success in the analysis of GAM and AGAM [6, 13–15]. The orbit of the spacecraft is studied around the center of mass of the system Sun-Mars. The Sun and Mars are assumed to be moving in circular Keplerian orbits around their common center of mass, and the spacecraft (third body with infinitesimal mass) is approaching Mars. In this case, the spacecraft has a drag device to increase the *A/m*, generating a larger effect of aerodynamic deceleration (*AD*). At the beginning of the simulations, the spacecraft is orbiting the Sun in a transfer orbit from the Earth, and then it makes a passage inside the atmosphere of Mars.

### 2.1 The AGAM

The dynamic equations in the plane of the rotating coordinate system in dimensionless variables, and restricted to the plane of the primaries, are derived from the potential of the CRTBP [13] with the addition of the acceleration due to drag, in a 2D maneuver. The two dynamics equations, as a function of the potential $\Omega$, are:

$$\ddot{x} = 2\dot{y} + \Omega_x + ADx \quad (1)$$
$$\ddot{y} = -2\dot{x} + \Omega_y + ADy \quad (2)$$

The potential is a function of the spacecraft´s position vectors to the Sun $\left(r_1 = \sqrt{(x+\mu)^2 + y^2}\right)$, to Mars $\left(r_2 = \sqrt{(x-1+\mu)^2 + y^2}\right)$ and to the gravitational constant relative to the mass of Mars ($\mu$). The potential function is:

$$\Omega = \frac{1}{2}(x^2 + y^2) + \frac{(1-\mu)}{r_1} + \frac{\mu}{r_2} \quad (3)$$

The derivation of the equations (1) to (3) are detailed in Ref. [13]. Using the variables of the system, the general form of the orbital energy of the spacecraft with respect to Mars is:

$$\varepsilon = \frac{V_2^2}{2} - \frac{\mu}{r_2} \quad (4)$$

### 2.2 Drag

The drag is a dissipative force acting in the direction opposite to the motion of the spacecraft relative to the atmosphere ($V_2$). The acceleration given by drag is a function of the *A/m*, the atmospheric density ($\rho$), and the drag coefficient ($C_D$), which is assumed to be constant (2.0 for the present scenario, which represents a flat plate in rarefied flow). The acceleration due to drag is described in more detail in [21], and it is represented by:

$$\overrightarrow{AD} = -\frac{C_D}{2}\left(\frac{A}{m}\right)\rho V_2 \overrightarrow{V_2} \quad (5)$$

For practical considerations, the constant value along the trajectory ($C_DA/m$) is known as the ballistic parameter ($B$).

Equation (6) shows the exponential model to calculate the density as a function of the altitude $h$, where $\rho_0$ is the atmospheric density at the surface of Mars (0.020 kg/m³) and $H$ is the scale height (11 000 m) [22]. The upper limit of the atmospheric model is assumed to be 2500 km of altitude.

$$\rho = \rho_0 e^{-\frac{h}{H}} \tag{6}$$

## 2.3 Perturbative Integral

This method measures the variation of the velocity of the spacecraft due to the drag during the whole period of the atmospheric passage. The *PI* is defined as the integral of the *AD*, from the beginning of the atmospheric passage ($t_i$) to the end of the passage inside the planetary atmosphere ($t_f$). In this case, it is selected the version of the *PI* that integrates the magnitude of the acceleration, because there is no compensation of accelerations coming from different directions during the trajectory, since the drag always removes energy from the spacecraft [5]. The study is made numerically, so the assumption of Keplerian orbits is not made. Equation (7) defines the *PI* used here.

$$PI = \int_{t_i}^{t_f} |\vec{AD}|\, dt \tag{7}$$

## 3. Parameters of the Simulations

Prograde and retrograde trajectories are used in the simulations made in the present paper. The pericenter altitude is selected to be inside the region where the atmosphere is dense enough to generate measurable perturbative effects, and the velocity at this point is calculated as the excess of velocity from a transfer orbit, to guarantee the approaching in hyperbolic conditions. To determine the initial conditions of the AGAM from the selected pericenter, the trajectories are propagated in negative times using the GAM, stopping when 0.5 DU from the center of Mars is reached. During this propagation, it is monitored the conservation of the Jacobi Constant with an accuracy lower than 1x10$^{-14}$ [6, 14, 15]. In the case of the AGAM, it is selected and A/m interval from 0.0 m²/kg to 20.0 m²/kg (the minimum value is used for the GAM and the maximum value represents the presence of solar sails or high drag devices). After one day of the passage of the spacecraft by the pericenter, the two-body energy relative to Mars and its distance is evaluated to classify the resulting trajectory.

The mathematical model was coded including the numerical integrator RKF-7/8 to solve the dynamic equations of motion (1) and (2). In the case of the equation (7), it was integrated numerically applying Simpson´s rule.

Due to the use of the dimensionless variables, the equivalent values of the non-dimensional units of the systems in the SI are presented in Table 1.

**Table 1 Equivalence of the Non-dimensional Units Sun-Mars to SI**

| Canonical Unit | Equivalence in SI |
|---|---|
| Distance Unit (1 DU) | 1.52367934 AU (2.27939186 x $10^{11}$ m) [24] |
| Velocity Unit (1 VU) | 24 131.229 m/s |
| Time Unit (1 TU) | 1 rad (2 623.8381442 hours) |
| Period of primaries | $2\pi$ rad (16 486.0613 hours) |
| Mars Radius (1 MR) | 3 397 200 m [24] |
| $\mu = \dfrac{m_{Mars}}{(m_{Sun} + m_{Mars})}$ | 3.227136860 x $10^{-7}$ |

## 4. Mapping the Resulting Orbits

The trajectories were propagated with *A/m* increasing in steps of 0.04 m²/kg to cover the effect of the deployment of the drag device. Initial perigee altitudes ($h_p$) change from 100 km to 150 km, in steps of 0.1 km.

For the trajectories arriving from the transfer orbit in GAM, two velocities at the pericenter of Mars were selected, 0.246 and 0.336 VU, which are calculated from a Hohmann transfers with perihelium equal to the Earth´s semimajor-axis and aphelion larger than the semimajor-axis of Mars, to generate a hyperbolic trajectory. The initial point is the same for all the trajectories (GAM and AGAM) since the trajectories come from the Earth. The angle of approach was selected to be 0°, indicating the pericenter in the apsis line, in front of the planet.

Figure 1 shows the three resulting trajectories classified in color scale, according to the value of the energy after the passage in prograde direction. White represents trajectories that, after the approach, ended in collisions with the planet. This region is increased for larger values of *A/m* and does not appear for the GAM (*B = 0.0 m²/kg*). Just above the collision region, it is in the capture zone (red color). After the capture, the continuum interaction of the spacecraft with the atmosphere makes them aerobraking trajectories. The capture region is the smallest of the three, because it depends on the specific combination between *A/m*, atmospheric density, and velocity, being sensitive to small density variations. The losses of energy in this region have intermediary values, which are large enough to transform the hyperbolic trajectory into an elliptic orbit, but not large enough to make a reentry in less than one day.

The last region is represented in blue. The trajectories located in this region are hyperbolic trajectories before and after the passage, which means that a successful AGAM occurred, and the gravity and the atmosphere of Mars was used to modify the interplanetary trajectory of the spacecraft.

Figures 1a and 1b show the resulting trajectories after the passage by the pericenter with a velocity of 0.246 VU. All the trajectories with *A/m* larger than 12 m²/kg ended in collisions in less than one day after the close approach (Fig. 1a). After 30 days of propagation, the white region increases and trajectories with A/m larger than 10.8 m²/kg also ended in collisions (Fig. 1b).

For 0.336 VU, the white and red regions are reduced, and the number of escapes increase, which indicates that the increase of the velocity generates higher values of drag, but reduces the duration of the passage, so reducing the effect of the perturbation (Fig. 1c and 1d). The differences observed between the left and right sides of Fig. 1 show the decrease in the escape altitude region and the reduction in the number of captured trajectories. The scenarios for 0.336 VU and altitudes larger than 140 km are ideal for the AGAM, when this maneuver is desired, like in the case of the returning of a spacecraft to the Earth, or to send it to other parts of the solar system.

The aerobraking maneuver is useful to place the spacecraft in a final or a parking orbit, keeping it in orbit for longer times with the reduction of A/m. In other words, the drag device is deployed for the capture and removed when it is desired to reduce the effects of the atmosphere, to stabilize the orbit of the spacecraft.

A total of 250000 trajectories were simulated for each scenario. For approaches with velocity of 0.246 VU, one day after the passage by the pericenter, only 5 536 trajectories resulted in aerocapture. After one month, this number was reduced to 943. For the scenario with 0.336 VU, 3 673 trajectories were captured after the passage, and one month later only 620 trajectories survived. It means that the complete mission must plan maneuvers to stabilize the orbit after insertion, but 30 days is a time long enough to do those maneuvers with all safety measures that are required. So, the aerobraking is an intermediate phase that reduces the energy of the trajectory of the spacecraft, reducing the fuel consumption for the complete maneuver.

The resulting orbits after one day of propagation are graphically represented in Fig. 2. In this case it was selected an A/m of 1.0 m$^2$/kg, at 140 km of h$_p$, after 1 day of the passage. Initial GA trajectories are represented in black, escape in blue, captures in red, and captures ending in collisions in green. Mars is represented by the orange circle and the atmospheric limits are shown by the blue dots. It is possible to observe the resulting geometry of the orbit after the closest approach. The trajectories are plotted in the plane of the synodic frame.

The second case analyzed here (or the third scenario) is the retrograde motion, which main effect is the increase in the relative velocity between the spacecraft and the atmosphere. In the rotational frame, the V$_P$ is 1.89 VU, then, in this case, the velocity relative to the atmosphere is maximized, reducing the collision regions, and increasing the number of escape trajectories, compared to the other scenarios. It was observed that only 796 orbits resulted in captures after one day of the approach, as shown in Fig. 3. In this case, for 30 days after the passage, all the captured trajectories continued in aerobraking, ending in collisions.

From Figs. 1 to 3, it is possible to conclude that the quantity of captures is maxima when the periapsis reference velocity for the GA is lower (in this case 0.246 VU). With the increase of the velocity at the periapsis of Mars, the capture region is reduced, both in the number of captures and in the altitude of the region, so moving the graphics to the right inferior corner, obtaining an increase in the escapes, and reducing the number of collisions. Then, one of the advantages of having larger velocities at the aphelion is the reduction of collisions, but it also reduces the area of the aerocapture region.

Figure 4 shows the limits of the capture region as a function of the altitude and the reference velocity at the pericenter of Mars, as calculated from an ideal GA in prograde direction. The increase of velocity reduces the collision region at the lowest altitudes, so increasing the escape region. The white region, between the red and blue regions, represents the capture trajectories ending in high eccentricity and large value of the semi-major axis, after 1 day of the passage. Then, these trajectories will collide with the planet in the passage by the pericenter.

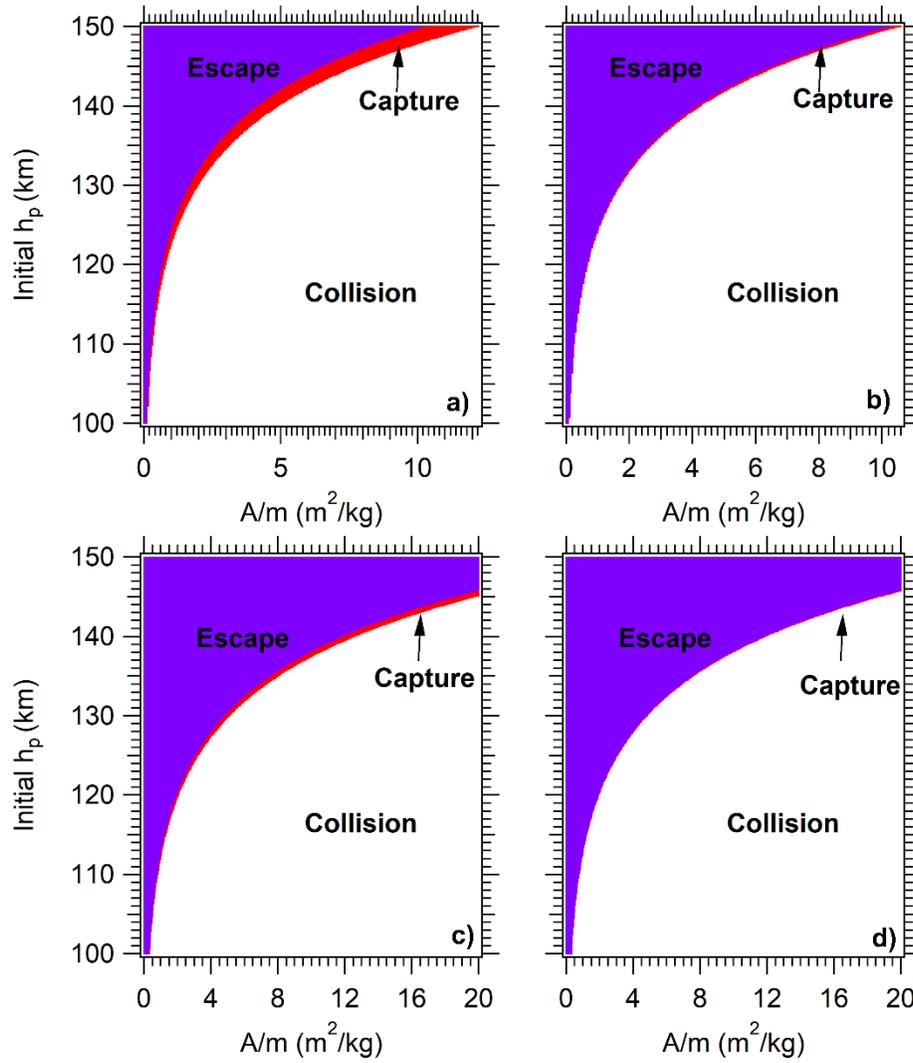

**Fig. 1 Final trajectories for $V_P$ = 0.246 VU (a, b), 0.336 (c, d); t = 1 (left), t = 30 (right) days after the passage.**

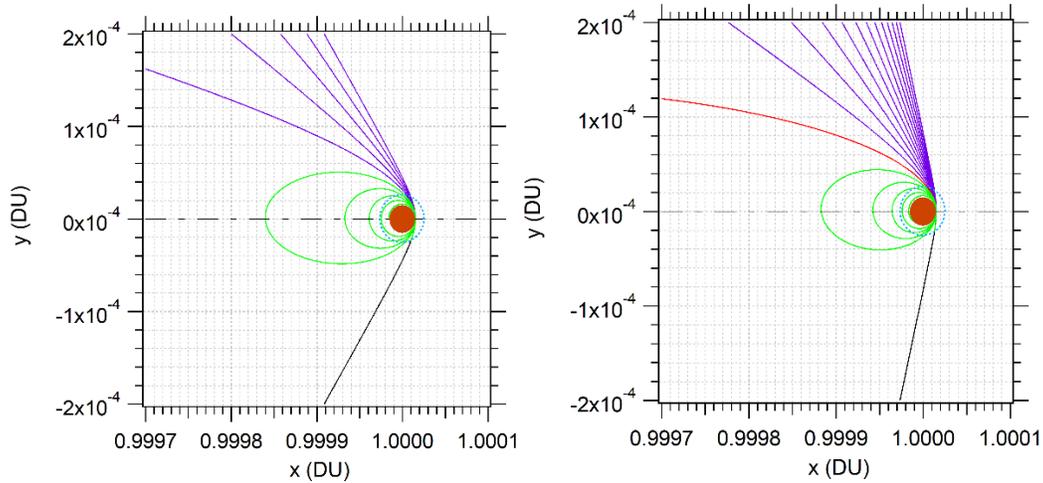

Fig. 2 Trajectories in the synodic frame with $V_P$ = 0.246 VU (left) and 0.336 VU (right).

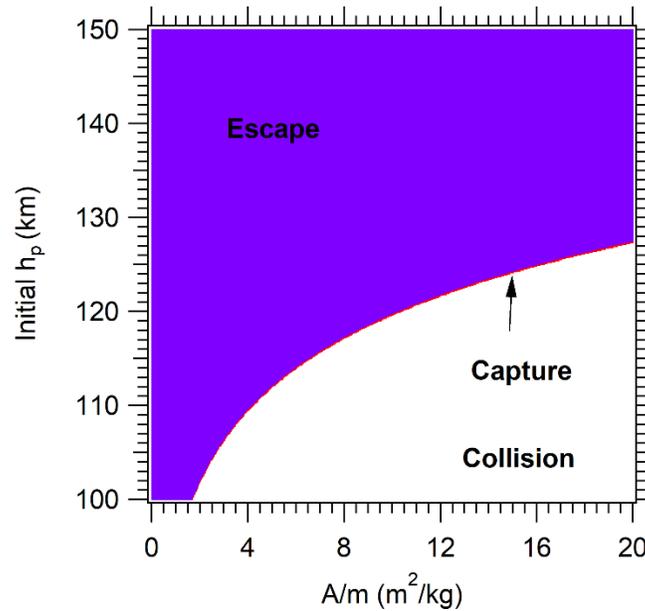

Fig. 3 Final trajectories for $V_P$ = 1.89 VU, retrograde direction.

The variations of the angle of approach for a selected altitude and velocity at periapsis during a GA, generate changes in the energy and velocity after the close approach [15]. In this case, it was analyzed the influence of the angle of approach in the collision, capture, and escape regions. The results show that the regions are constant for a selected velocity of approach and *A/m*, independent of the angle of approach since the relative velocity to the atmosphere is the same.

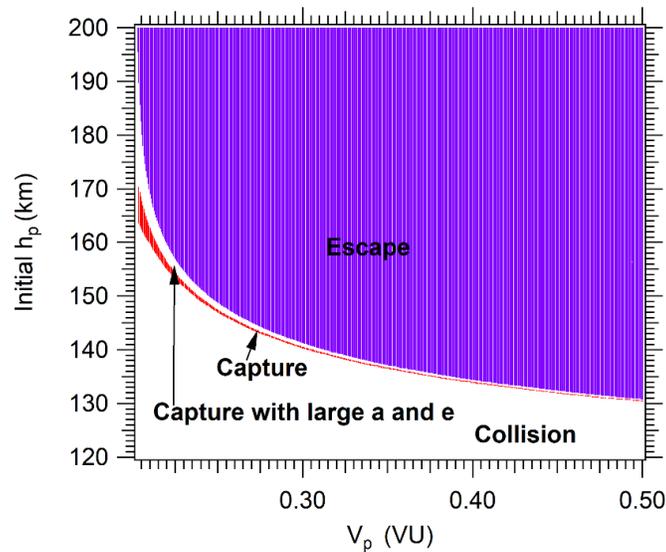

**Fig. 4 Final trajectories for A/m = 10.0 m²/kg, t = 1 day after the passage.**

## 5. Mapping the Resulting Orbits

The resulting trajectories after the closest approach (aerocapture and AGAM), are analyzed using the PI technique, which means that we will measure the value of the total variation of velocity (DV) given by the drag. This method can measure the reduction in costs of fuel obtained by the AGAM, because the PI measures the contribution of the atmosphere to the maneuver, which is equivalent to the impulse necessary to reduce the velocity of the spacecraft to place it in its final orbit.

With the use of the PI maps, it is possible to find the existence of a lower limit of 1.0 km/s and an upper limit of 2.05 km/s for the captured orbits (velocity of the GAM at periapsis is 0.246 VU). Orbits with PI lower than 1.0 km/s resulted in escape trajectories, while values larger than 2.05 km/s generate collisions (see Figs. 5 and 6). For the GAM of with 0.336 VU, the captures occur shows PI values larger than 3.2 km/s and lower than 4.4 km/s (Figs. 5 and 6, right sides). A larger difference is observed in the boundary regions between captured and escaped orbits, because the captured orbits maintain the energy losses due to the constant interaction with the atmosphere, unlike the escape orbits, that interact in a short interval of time with the atmosphere and then leave the proximity of Mars quickly. Figure 5 shows the strong reduction of DV for higher altitudes and lower values of *A/m*, going to zero for the lowest values of *A/m*. This is the first time that an integral index is used to quantify the aerodynamic contribution of the AGAM on Mars, and the results showed that it generates interesting general maps that can be used for mission designers when planning the capacity of the propulsion system or DV budget for the mission. In particular, the PI measures accurately how much the atmosphere reduces the DV for larger altitudes and lowest values of A/m, with the results described by non-homogeneous thin layers. The results of the Figs. 5 and 6 were propagated in the prograde direction, for 1 day after the approach.

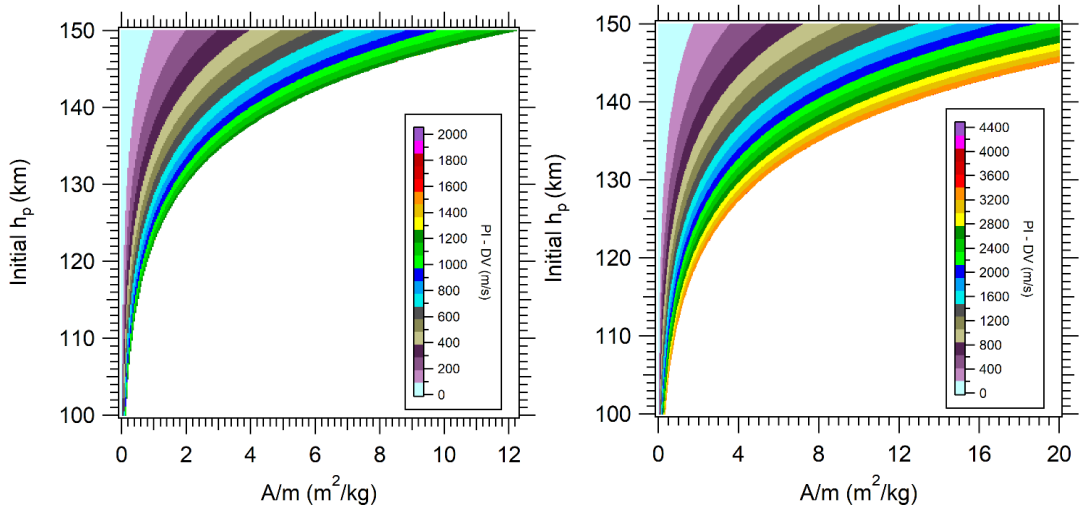

**Fig. 5 PI maps for V$_P$ = 0.246 VU (left), 0.336 VU (right).**

The PI maps for captured orbits are presented in detail in Fig. 6, which shows in more detail the boundary layers of maximum DV before the collision and the minimum DV before the escape region. The layer near the collision zone, for 0.246 VU, indicates a DV around 1.3 km/s, decreasing in the direction of the escape zone. When the velocity at periapsis is 0.336 VU, the layer has values from 3.4 km/s to 3.2 km/s. The color scale in the figures presents higher values, which represents a few trajectories that result in this condition after the capture.

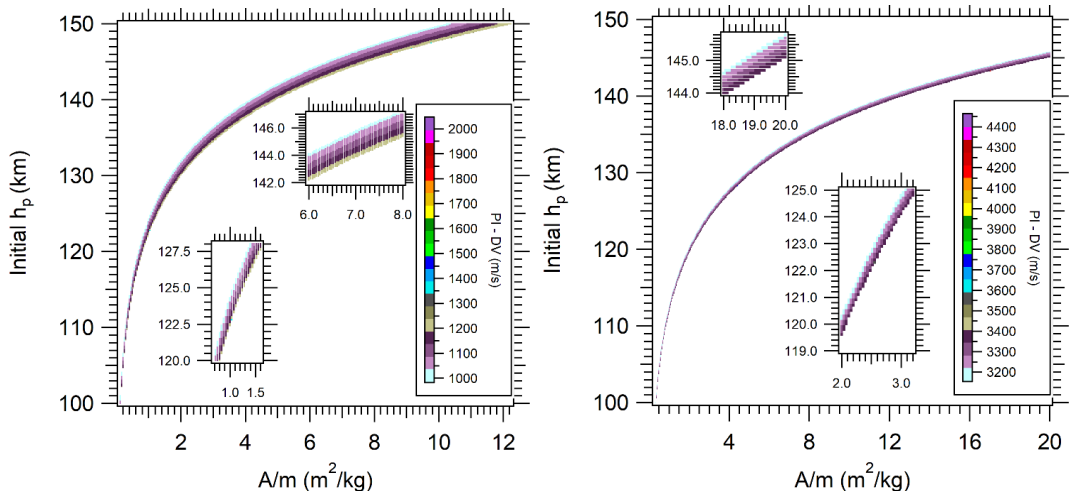

**Fig. 6 PI maps for V$_P$ = 0.246 VU (left), 0.336 VU (right).**

In the case of retrograde trajectories, due to the higher values of the velocity (the drag force increases in magnitude with the square of the velocity) therefore, it generates larger values of DV (see Fig. 7). It is also shown that only trajectories with variations between 38 and 37 km/s are captured. A large amount of energy is dissipated due to the interaction with the atmosphere. Figure 7 presents the PI maps for retrograde approaches after 1 day.

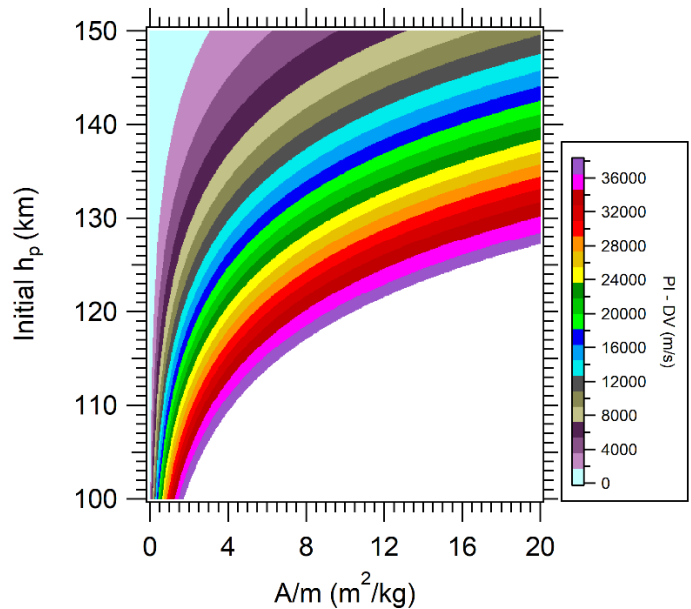

**Fig. 7 PI maps for VP= 1.89 VU.**

With the goal of comparing the results given by the PI-DV with the single impulse maneuver (without drag), it was determined the eccentricity and semi-major axis of the resulting trajectories for one day after the aerocapture. From the initial trajectory, it is possible to calculate the DV required. For example, in the case of a capture with *A/m* = 1.0, at 122.4 km of pericenter altitude and velocity 0.246 VU at periapsis, the semi-major axis resulted from the aerocapture is 6.37819 MR with a PI-DV of 1207.33 m/s. At this point, it is calculated the orbital velocity of the elliptical orbit that the spacecraft would have if the atmosphere were not present. The required DV to transform the orbit, assumed to be the same final orbit given by the maneuver using the atmosphere, can be calculated from the Vis viva equation, and the result is 1115.4 m/s. This is the value saved in the propulsion system by the effect of the atmosphere. Table 2 shows the results of the PI-DV given by drag and the DV required to capture the spacecraft (in GA) and to place it in the same orbit that the atmosphere did. In this case, the values of the impulses are lower than the results obtained using the PI-DV, because the impulses are instantaneous and do not consider the effects of the continuous dissipation from the atmosphere, before the passage by the pericenter.

The use of the PI also allows the identification of the variations in the trajectories. In the case of aerocaptures, the PI shows an initial large variation in a short time, showing the DV dissipated by the atmosphere during the first approach. In the next days, the variations in DV are negligible compared to the values required for the capture but showing small variations that represents the aerobraking. After several days, a largest variation occurs, which represents the change from decay orbit condition to reentry and collision. Then, the PI-DV as a function of time is useful to observe these behaviors, to quantify the DV dissipated by the atmosphere and the duration of the changes. In Fig. 8 it is presented the evolution of the PI-DV for trajectories in prograde direction with 10.0 $m^2$/kg and initial velocity at the pericenter of 0.246 VU.

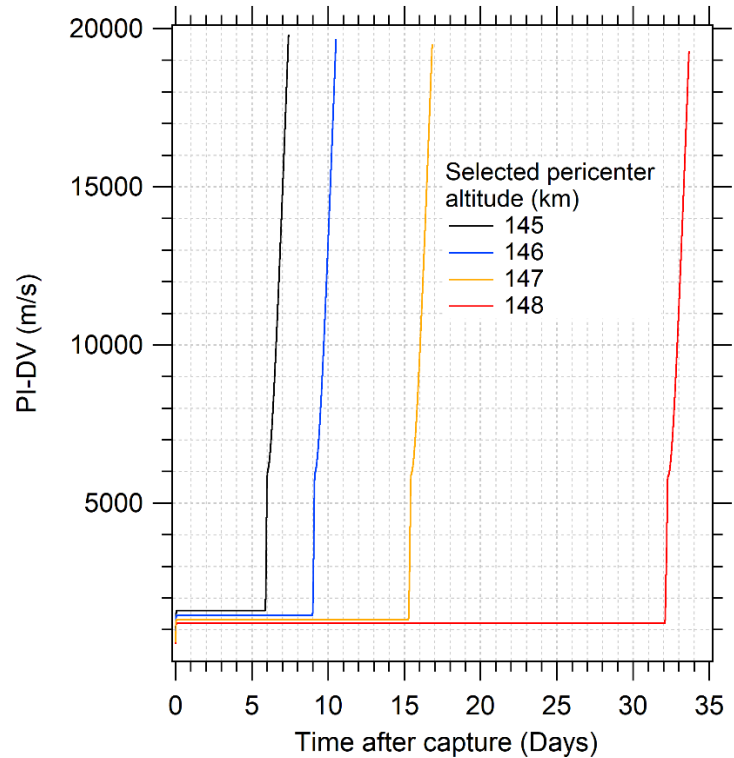

**Fig. 8 PI-DV as a function of time.**

**Table 2. DV comparison between a pure impulsive maneuver and AGAM**

| A/m (m²/kg) | $V_p$ (VU) | $h_p$ (km) | a (MR) | PI-DV (m/s) | Impulse DV (m/s) |
|---|---|---|---|---|---|
| 1.0 | 0.246 | 122.4 | 6.37819 | 1207.33 | 1115.40 |
| 11.0 | 0.246 | 150.0 | 16.278 | 1101.48 | 992.29 |
| 1.0 | 0.336 | 112.0 | 7.41537 | 3343.00 | 3258.63 |
| 11.0 | 0.336 | 139.0 | 24.0415 | 3240.32 | 3138.87 |

**Conclusions**

In this research, for the first time, it was applied the technique known as PI to measure the total impulse obtained from the Martian atmosphere during the close approach. The PI technique allows us to measure the savings that the atmosphere provided during aerocapture, aerobraking, and AGAM. The results show a particularly good agreement between the PI and the savings calculated by the analytical impulsive method. It means that the PI is valid to be used to predict the savings generated by drag. This technique should be used to complement the design of future aeromaneuvers.

With the use of the PI-DV technique, it was possible to map the regions of capture, as well as to calculate the DV dissipated by the atmosphere, the savings obtained for the propulsion system, and to quantify the boundary values of the capture regions. Therefore, we can also see the best form to use the atmosphere of Mars for this maneuver and to quantify accurate the savings obtained.

**Acknowledgments**

The authors wish to express their appreciation for the support provided by Science and Technology Institute (ICT – UNIFESP) and by the National Institute for Space Research (INPE).

This work was supported by grants # 2019/26605-2 and # 2016/24561-0, from São Paulo Research Foundation (FAPESP); grants # 406841/2016-0, 301338/2016-7 and 303102/2019-5 from the National Council for Scientific and Technological Development (CNPq); grant # 88882.317514/2013-01 from the National Council for the Improvement of Higher Education (CAPES).

**Authors' contributions**

The authors: Jhonathan Murcia Piñeros, Antonio F.B.A. Prado, Walter A. dos Santos e Rodolfo Vilhena de Moraes contributed to the study conception and design. The authors performed material preparation, data collection and analysis. All authors read and approved the submitted version.

**Competing interest**

The authors declare that they have no conflicts of interest with research institutions, professionals, researchers and/or financial supports.

**Data availability**

All data generated or analyzed during this study are included in this published article in form of figures.

**Ethical approval**

The submitted work is original and not have been published elsewhere in any form or language. The work presents the results of a single study. The results are presented clearly, honestly, and without fabrication, falsification, or inappropriate data manipulation. No data, text, or theories by others are presented as if they were the author's own